\documentclass[12pt]{article}
\usepackage{graphicx,amssymb,amsmath,amsthm,cite}

\newtheorem{proposition}{Proposition}

\def\be{\begin{equation}}
\def\ee{\end{equation}}
\def\ben{\begin{eqnarray}}
\def\een{\end{eqnarray}}

\newcommand{\la}{\langle}
\newcommand{\ra}{\rangle}

\newcommand{\til}{\tilde}

\newcommand{\kfV}{|f_{\SV} \ra}

\newcommand{\kfWC}{|f_{\SWC} \ra}

\newcommand{\kvi}{|v_i\ra}
\newcommand{\bvi}{\la v_i |}

\newcommand{\bwi}{\la w_i |}
\newcommand{\bui}{\la u_i |}
\newcommand{\kui}{|u_i\ra}

\def\op{\hat{P}}

\def\SS{{\cal{S}}}

\def\SV{{\cal{V}}}
\def\SW{{\cal{W}}}

\def\SWr{\til{{\cal{W}}}_r}
\def\SWo{\til{{\cal{W}}}_0}
\def\SVr{\til{{\cal{V}}}_r}
\def\SVo{\til{{\cal{V}}}_0}

\def\SVKF{{\cal{V}}_K}

\def\SW{{\cal{W}}}

\def\SWC{{\cal{W}^\bot}}
\def\SVC{{\cal{V}^\bot}}
\def\EVW{\hat{E}_{\SV \SWC}}
\def\EVWr{\hat{{E}}_{\SVr \SWr}}
\def\EVV{\hat{E}_{\SV \SVC}}

\newcommand{\Spann}{\text{span}}

\def\C{\mathbb{C}}
\def\emptyy{\{0\}}

\title{Nonlinear non-extensive approach for identification 
of structured information}
\author{Laura Rebollo-Neira\\
Mathematics, Aston University\\ 
Birmingham, B4 7ET, UK\\
A. Plastino\\
IFLP-CCT-Conicet\\
Universidad National University La Plata\\
CC 727, 1900 La Plata, Argentina}

\begin{document}
\maketitle

\begin{abstract}
\vspace{1cm}
The problem of separating structured information representing
phenomena of differing natures is considered.
A structure is assumed to be independent of the
others if can be represented in a complementary subspace.
When the concomitant subspaces are well separated the problem 
is readily solvable by a linear technique. 
Otherwise, the linear approach fails to correctly discriminate 
the required information. Hence, a non extensive approach
is proposed. The resulting nonlinear technique is  shown 
to be suitable for dealing with cases that cannot be tackled by
the linear one.

\end{abstract}

\section{Introduction}
We consider the problem of discriminating information produced 
by phenomena of differing natures, via inverse methods. This involves  
the study of the physical state of a  system by analyzing its 
response to some external interaction. 
We refer to the interactive carrier as {\em {input 
signal}} and to the system's reaction  as {\em{signal response}}. 
Unfortunately, a particular response is not 
always directly available, as one may receive it `disguised' 
by the interference with another independent 
phenomenon not being the focus of specific interest. 
In this paper we restrict our consideration to responses evoked 
by statistical systems.  By this we understand 
 systems which are fully 
characterized by of a probability distribution indicating 
either the population of subsystems compressing the whole system, or 
the degree of uncertainty about the system being in one of its possible 
states. We regard both situations to be identical in the
description and refer to subsystems as system's states.

In order to formulate the problem 
let us use the label `$i$', ranging from 1 to $M$, 
to denote the $i$-th state of a system which is characterized by a 
probability $p_i$. Adopting Dirac's notation we indicate by a ket
$|f_\SV\ra$ the system's response to some input signal 
and by $|v_i\ra$ the corresponding response of the $i$-th state. 
Consequently, the system's signal response satisfies
$$|f_\SV\ra \propto \sum_{i=1}^M p_i |v_i\ra.$$
This equation is transformed into an equality by simply
relaxing the condition $\sum_{i=1}^M p_i=1$, so that 
$$|f_{\SV} \ra = \sum_{i=1}^M c_i |v_i\ra,$$
where the coefficients in the superposition 
are not necessarily normalized to unity.
As already stated, we are interested in the problem of 
discriminating $|f_\SV\ra$ from a given signal $|f\ra$
of which $|f_\SV\ra$ is a component. 
Out of the many situations involving this problem 
it immediately comes to our mind the intensity of X-rays
produced simultaneously by dispersion and diffraction
or an infrared emission spectrum superimposed
to blackbody radiation. In order to model all relevant cases 
we assume that, rather than
$\kfV$, the available signal is $|f\ra= \kfV + \kfWC$, 
where $\kfWC$ is produced by an independent phenomenon. 
We focus on those cases ensuring a unique decomposition, i.e., 
we further assume that the subspaces hosting the components $\kfV $
and $\kfWC$ are complementary. However, the focus of 
our interest refers to complementary subspaces being 
close enough together to move the problem of separating the components 
far away from the trivial one. 
Certainly, if the subspaces  hosting the signal components  are
well separated, the problem  is 
readily solvable by means of an oblique projection onto one of 
the subspaces and along the other \cite{BS94,Reb07a}. 
Contrarily, if the subspaces are 
not well separated the construction of the necessary projector 
becomes ill posed and the problem needs to 
be tackled in an alternative way. In this Communication we 
address the matter by including a hypothesis upon the system 
producing the signal response. {\em{We assume that the population 
of states is $K$-sparse in the sense that, out of the $M$ possible 
states of the system, only $K<M$ of them 
are characterized by a significant  probability.}} 
Nevertheless, the hypothesis generates, in general, an intractable 
problem, because of course the populated states are unknown and 
the number of possibilities of populating $K$ states out of 
$M$ is a combinatorial number $\tbinom{M}{K}$. This makes  
the exhaustive search for the unknown states an impossible task 
for most values of $M$ and $K$.  In recent
publications \cite{Reb07b,Reb09} a greedy strategy for making  
the search tractable has been proposed. In the 
present context, the proposal 
of that publications  implies to assume 
a priori that no state is populated 
and looks for the populated ones in a stepwise manner.
Here we investigate  the possibility of addressing 
the problem from the opposite view point. Assuming
a priori that all the states are equally populated, 
we will determine the actual population of each state via the minimization 
of the $q-$norm like quantity
$\sum_{i=1}^M |c_i|^q,\quad 0<q\le 1.$
The  minimization  of this quantity as an appropriate  criterion for
 determining a sparse solution to an under-determined linear 
 system is  discussed in \cite{Wic94,RD99}.
For nonnegative and normalized to unity coefficients $c_i,i=1,\ldots,M$, 
this quantity is closely related to  
 the non-extensive entropic measure
broadly applied in physics \cite{Tsa88,Tsa09,Pla1,Pla2,LS00,AO01} since
Tsallis introduced it as the essential
ingredient of his thermodynamic analysis framework \cite{Tsa88}.
In the present context the value of $q$ plays a particular role.  
By choosing $0<q \le 1$ we introduce an assumption on the  sought 
distribution. {\em{ We assume that not all 
the possible states in a system's model are significantly populated}}. 
This assumption is meant to compensate for the actual overestimation 
of possibilities one usually makes when a system's 
signal response is modelled mathematically. 

The paper is organized as follows: Section \ref{sec2} 
introduces the mathematical setting of the problem  
and discusses the construction of oblique projectors.
 Section \ref{sec3}
 remarks the need for nonlinear approaches to
 separate signal components living in 
 subspaces which are `theoretically' complementary,  
 but close enough to prevent the components discrimination being 
 realized by a linear operation. 
 The proposed strategy, based on the minimization of 
 the  $q$-norm$^q$  $\sum_{i=1}^M |c_i|^q$,    
 subject to recursively  selected constraints, 
 is discussed in Section \ref{sec4} and illustrated  in 
 the same section by a numerical simulation. The 
  numerical experiment is especially  designed to  
  highlight the robustness of the proposed approach against 
 significant error in the data. 
The conclusions are presented in Section \ref{conclu}. 
                                        
\section{Mathematical setting of the problem}
\label{sec2}
As already mentioned, adopting Dirac's notation we represent 
the response of a statistical system to some external interaction as
 $|f_\SV\ra$, which is expressible in the form
\be
\label{mod}
|f_{\SV} \ra = \sum_{i=1}^M c_i |v_i\ra.
\ee
Since the kets are elements of an inner product space, their square norm 
is induced by the inner product, i.e., $\|\kfV \|^2 =\la f_{\SV} \kfV$. 

The problem we are concerned with entails to `rescue' a 
ket response $\kfV$ from an available mixture $|f\ra=\kfV +\kfWC$, 
where $\kfWC$ is produced by an independent phenomenon (e.g. 
a structured interference that one would call {\em{background}}
referring to a persistent effect out of the focus of the main interest).

Denoting $\SV= \Spann\{\kvi\}_{i=1}^M$ and assuming that the 
subspace $\SWC$ such that $\kfWC \in \SWC$ is known, we 
restrict considerations to the  case $\SV \cap \SWC = \emptyy$ so as 
to ensure the uniqueness of the decomposition $|f\ra=\kfV +\kfWC$. 
Such a problem has a straightforward `theoretical' solution. 
Certainly, from $\hat{E}_{\SV \SWC}$, the oblique projector 
onto $\SV$ along $\SWC$, one immediately has 
$$\hat{E}_{\SV \SWC} |f\ra = \hat{E}_{\SV \SWC} (\kfV + \kfWC)= \kfV.$$
However, as will be discussed in the next section, when the 
subspaces $\SV$ and $\SWC$ are not well separated the numerical 
construction of $\hat{E}_{\SV \SWC}$ becomes ill posed,  thus preventing 
the signal separation to be correctly realized.
\subsection{Construction of Oblique Projections}
Let us recall that every idempotent operator is a projector. 
Hence, an operator $\hat{E}$ is a projector if
$\hat{E}^2 = \hat{E}$. The projection is
along its null space and onto its range.
When these subspaces
are orthogonal $\hat{E}$ is called an
orthogonal projector, and it is the case if and only if
$\hat{E}$ is self-adjoint. 
Otherwise it is called an {\em {oblique projector}}. For a 
 good and amusing introduction to oblique projectors in the context of 
 signal processing we refer to 
\cite{Eld03} and for advanced
theoretical study of oblique projector operators in infinite dimensional 
spaces to \cite{CMS05a,CMS05b}. Here we will restrict ourselves 
to issues related to numerical constructions.

Assuming that $\SV \cap \SWC = \emptyy$ 
the oblique projector operator onto $\SV$
along $\SWC$ will be represented as above.
Then $\EVW$ satisfies
$\EVW^2= \EVW$
and, consequently,
\ben
\EVW |g \ra &=&  |g \ra, \quad \text{if} \quad |g \ra \in \SV  \nonumber\\
\EVW |g \ra&=& 0, \quad \text{if} \quad  |g \ra \in \SWC. \nonumber
\een
In the particular case for which $\SWC =\SV^\bot$ the operator
$\EVV$ is an orthogonal projection onto $\SV$. For indicating  an orthogonal
projector onto a subspace, ${\cal{X}}$ say, we  use the
particular notation $\op_{\cal{X}}.$ 

Given $\SV$ and $\SWC$, in oder to construct $\EVW$ 
we define $\SS$ as the direct
sum of $\SV$ and $\SWC$, which we express as
$$\SS= \SV \oplus \SWC.$$
Let $\SW= (\SWC)^\bot$ be the orthogonal complement of $\SWC$ in $\SS$. Thus
we have
$\SS= \SV \oplus \SWC=\SW \oplus^\bot \SWC,$
where the operation $\oplus^\bot$ indicates the orthogonal sum referring to 
the direct sum of orthogonal subspaces.
Assuming that a set   
$\{y_j\}_{j=1}^{J}$ spanning $\SWC$ is known, we can always construct 
the orthogonal projector $\op_{\SWC}$  to be expressed in the form
$$\op_{\SWC} = \sum_{j=1}^{J'} |o_j\ra \la o_j |,\,J'\le J,$$
where vectors  $\{|o_j\ra\}_{j=1}^{J'}$ span $\SWC$ and are orthonormal, 
while the given set $\{|y_i\ra\}_{j=1}^{J}$ is not necessarily orthogonal, nor
even linearly independent.

From the set $\{\kvi\}_{i=1}^M$, spanning $\SV$, 
a spanning set for $\SW$ is readily obtained as
\be
\label{ku}
\kui=\kvi - \op_{\SWC} \kvi = \op_{\SW}\kvi,\,i=1,\ldots,M.
\ee
Denoting by $\{| i \ra \}_{i=1}^M$ the standard orthonormal basis for
$\C^M$,
operators $\hat{V}: \C^M \to \SV$ and $\hat{U}:\C^M \to \SW$ are defined 
as
$$\hat{V}=\sum_{i=1}^M \kvi  \la i|,
\;\;\;\;\;\;\;\;\;\;\;\;\;\;
\hat{U}=\sum_{i=1}^M  \kui   \la i|.$$
Consequently, the adjoint operators $\hat{U}^\ast$ and
$\hat{V} ^\ast$ are 
$$\hat{V}^\ast=\sum_{i=1}^M  |i \ra \bvi,
\;\;\;\;\;\;\;\;\;\;\;\;\;\
\hat{U}^\ast=\sum_{i=1}^M  |i \ra  \bui.$$
Since
$\op_{\SW} \hat{V}= \hat{U}$ and
$\hat{U}^\ast \op_{\SW} = \hat{U}^\ast$,
 the operator $\hat{G}: \C^M \to \C^M $ given below 
$$\hat{G}=\hat{U}^\ast  \hat{V}= \hat{U}^\ast  \hat{U}$$
is a self-adjoint operator. The elements of its matrix representation
are 
$$\la i| \hat{G} | j\ra= \la u_i |v_j\ra= \la  u_i |u_j\ra, \, i,j=1,\dots,M.$$
In terms of the above defined operators the oblique 
projector $\EVW$ is expressed as  
\be
\label{yoni}
\EVW= \hat{V} \hat{G}^\dagger \hat{U}^\ast
\ee
or, equivalently, 
\be
\label{obli2}
\EVW= \sum_{i=1}^M \kvi \bwi,
\ee
where
\be
\bwi=  \la i| \hat{G}^\dagger  \hat{U}^\ast= \sum_{j=1}^M 
   \la i| \hat{G}^\dagger |j \ra \la u_j| = \sum_{j=1}^M
   {g}^\dagger_{i,j} \la u_j|.
\ee
with ${g}^\dagger_{i,j}=  \la i| \hat{G}^\dagger |j \ra$ the 
 element $(i,j)$ of a matrix $G^\dagger$ indicating the 
pseudo inverse of $\hat{G}$.
It is actually straightforward to verify that $\EVW$ given in
(\ref{obli2}) satisfies the required properties. Namely,
i)$\EVW^2 = \EVW$, ii)
$\EVW \kfV = \kfV$, for all $\kfV \in \SV$, and
iii)$\EVW | g  \ra = 0$ or all $ | g \ra  \in \SWC.$

{\bf{Note}:}
The condition $\SV \cap \SWC = \emptyy$ implies that the dimension of $\SV$
is equal to the dimension of $\SW$. Accordingly, if the
spanning set $\{\kvi\}_{i=1}^M$ is linearly independent, operator $\hat{G}$
has an inverse. Nevertheless, the independence of $\{\kvi\}_{i=1}^M$
is not required, so that an inverse for $\hat{G}$ need not exist. For
the sake of generality we use $\hat{G}^\dagger$, which it is equal to 
$\hat{G}^{-1}$ when such an inverse does exist.

Let us stress that, since operators $\hat{V}$ and $\hat{U}$ are given in 
terms of spanning sets for the spaces
$\SV$ and $\SW$, respectively, {\em any} such spanning sets
can be used. This possibility yields a number of 
different ways of computing
$\EVW$, all of them, of course, theoretically equivalent but not necessarily 
numerically equivalent when the problem is ill posed. 

Considering that $|\psi_n \ra \in \C^M, \,n=1,\ldots,M$,
are the eigenvectors of $\hat{G}$
and assuming that there exist $N$ nonzero eigenvalues
$\lambda_n,\,n=1,\ldots,N$, on taking these eigenvalues
in descending order we can express the matrix elements of the
Moore-Penrose pseudo inverse of $\hat{G}$ 
as:
\be
\hat{G}^\dagger = \sum_{n=1}^N |\psi_n \ra \frac{1}{\lambda_n} \la \psi_n|.
\ee
Moreover, the orthonormal vectors
\be
\label{xi}
|\xi_n\ra= \frac{\hat{U}|\psi_n\ra }{\sigma_n},\quad \sigma_n=\sqrt{\lambda_n},\quad
n=1,\ldots,N
\ee
are singular vectors of $\hat{U^\ast}$, which satisfy
$\hat{U}^\ast |\xi_n\ra =\sigma_n |\psi_n\ra$, as it is 
immediate to verify.
By defining now the vectors $|\eta_n\ra,\,n=1,\ldots,N$ as
\be
\label{eta}
|\eta_n\ra =  \frac{\hat{V}{|\psi_n\ra}}{\sigma_n},
\quad \,n=1,\ldots,N,
\ee
the projector $\EVW$ in \eqref{obli2} is recast as
\be
\label{evw2}
\EVW=\sum_{n=1}^N |\eta_n \ra \la \xi_n |.
\ee
\begin{proposition}
\label{bio}
The vectors $|\xi_n\ra \in \SW,\,n=1,\ldots,N$ and
$|\eta_n\ra \in \SV,\,n=1,\ldots,N$
given in \eqref{xi} and \eqref{eta}
are biorthogonal to each other and span $\SW$ and $\SV$, respectively.
\end{proposition}
The proof of the above proposition is given in Appendix A.

It is immediate to verify that the representation 
\eqref{obli2} of  $\EVW$  also arises from
\eqref{evw2}, since
\be
\label{wdu2}
|w_i\ra = \sum_{n=1}^N  |\xi_n \ra \frac{1} {\sigma_n} \la  \psi_n|i\ra,
\quad i=1,\ldots,M.
\ee
Moreover, the representation \eqref{obli2} can be numerically realized 
in different ways by using different spanning sets to 
compute the operator $\hat{G}$. Indeed, by orthogonalizing 
$\{|u_i\ra\}_{i=1}^M$ to obtain the orthogonal set $\{|q_i\ra\}_{i=1}^{M'},
\, M'\le M$
also spanning $\SW$, the matrix elements of 
operator $\hat{G}_q$
(replacing $\hat{G}$) are
 given  as $\la q_i |v_j\ra$ (or equivalently  as
$\la q_i |u_j\ra)\, i,j=1,\ldots,M$. Thus, 
vectors $\la w_i|$ in 
\eqref{obli2} are calculated as 
$\bwi = \sum_{j=1}^{M'}\la i |\hat{G_q}^\dagger |j \ra \la q_j|$.

If spaces $\SV$ and $\SWC$ are not too close, 
which is reflected by  the fact that the 
non zero singular values of $\hat{U}$ are not too small, all the 
constructions of $\EVW$ are equivalent. However, as will be discussed 
in the next section, the existence of small singular values may  
render all the possible  numerical constructions of $\EVW$ incapable
of producing the expected signal splitting by the operation $\EVW |f\ra$. 
\section{The need for non-linear approaches}
\label{sec3}
This section is dedicated to illustrating, by recourse to
a numerical example, the crucial role that nonlinear approaches 
could play for the success of discriminating signal components when 
the concomitant linear problem is ill posed. 

{\bf{Numerical Example}} Let $\SV$ be the cardinal cubic spline
space with distance $0.01$
between consecutive knots, on the interval $[0,1]$. This is
a subspace of dimension $M=103$, which we span using a
B-spline basis $$B=\{B_i(x),\,x\in [0,1]\}_{i=1}^{103}.$$
The functions $B_i(x)$ in $B$  are obtained by 
translations of a prototype function 
and the restriction to the interval $[0,1]$ \cite{Sch81,AR05}. A 
few of such functions  
are plotted in the left hand graph of Figure 1. 
Here the inner product is defined  as
$\la f | h\ra=\int_{0}^{1} f(x)^\ast h(x)\,dx$, and all
the integrals are computed numerically.

\begin{figure}[!ht]
\label{f1}
\begin{center}
\includegraphics[width=8cm]{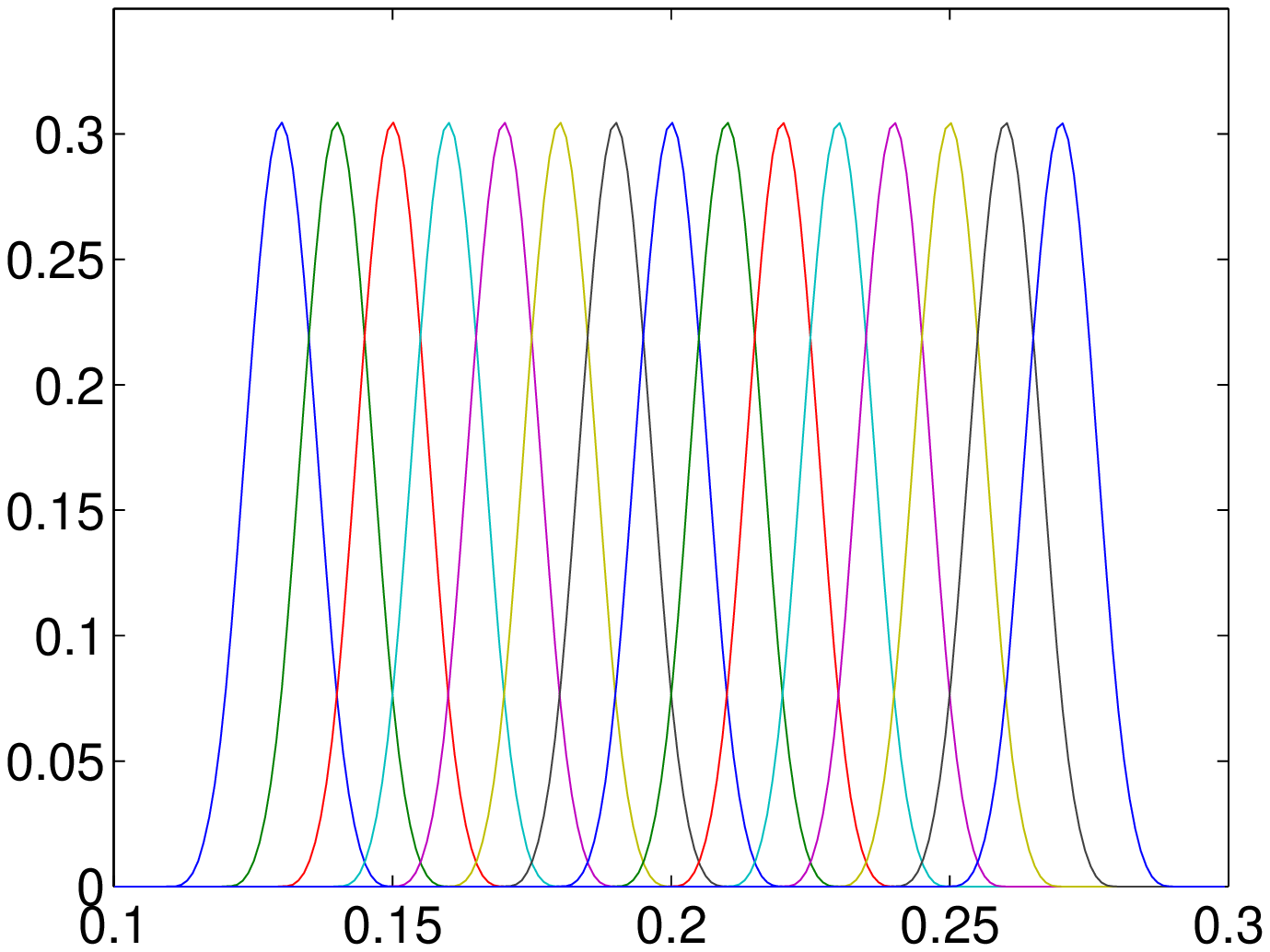}
\includegraphics[width=8cm]{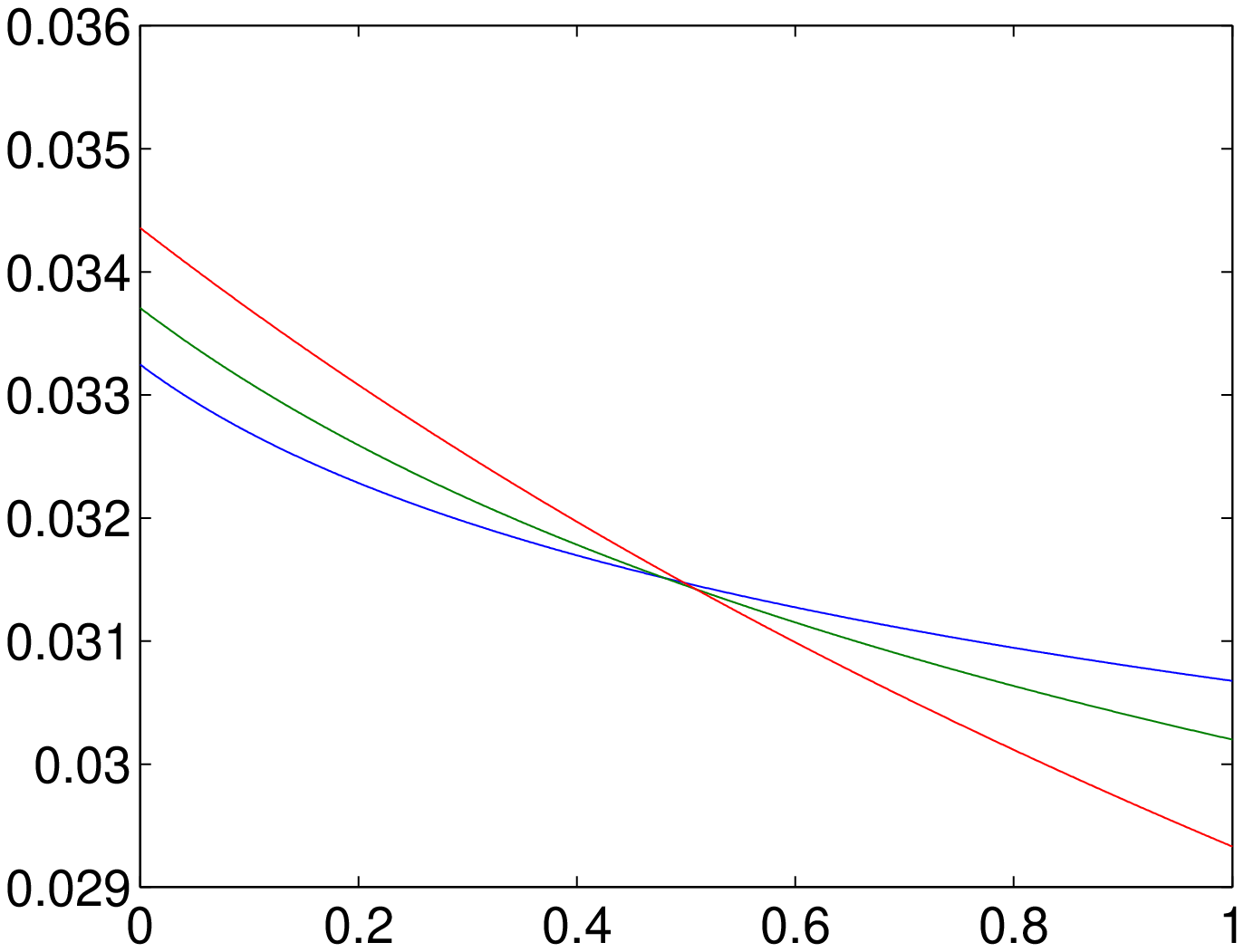}
\end{center}
\caption{\small{Left graph: cubic B spline functions, in the rage $x 
\in [0.1, 0.3]$,
from the set spanning the space of the signal response.
Right graph: three of the functions spanning the space of the 
background.}}
\end{figure}

Randomly taking 30 B-splines $\{B_{\ell_i}\}_{i=1}^{30}$ from 
$B$ we simulate a spectrum by a weighted superposition of such functions, i.e., 
the response signal is modelled as 
\be
\label{sir}
\la x \kfV =f_\SV(x)=\sum_{i=1}^{30} c_{\ell_i} B_{\ell_i}(x),\quad x\in [0,1],
\ee
with the coefficients $c_{\ell_i}$ randomly chosen from 
$[0,1]$. 

We simulate a background  by considering that it belongs 
to the subspace $\SWC$ spanned by the set of functions
$$Y=\{y_j(x)=(x+0.01j)^{-0.01j},\,x\in [0, 1]\}_{j=1}^{50}.$$
A few functions from this set are plotted in the  right hand 
graph of Figure 1 (normalized to unity on $[0, 1])$.
The background, $g(x)$, is generated by 
the linear combination 
\be
\label{bac}
\la x | g \ra = g(x)= \sum_{j=1}^{50} j^4e^{-0.05j}y_j(x).
\ee
To simulate the data 
we have perturbed the superposition of $\eqref{sir}$ and $\eqref{bac}$,  by 
`very small' Gaussian errors (of variance up to $0.00001\%$ the value 
of each data point) 
and plotted the  simulated data in the left hand graph of 
Figure 2. 

This example illustrates well  how sensitive to
errors the oblique projection is.
The subspaces we are dealing with are disjoint: the
last five singular values of operator $\hat{U}^\ast$ (c.f. \eqref{xi})
are:
$$0.3277,  0.3276, 1.0488 \times 10^{-4},
 6.9356 \times 10^{-8}, 2.3367 \times 10^{-10},$$
while the first is $\sigma_1=1.4493$.
The smallest singular value cannot be considered a
numerical representation of zero when the calculations are being
carried out in double precision arithmetic. Hence, one can assert that
the condition $\SV \cap \SWC = \emptyy$ is fulfilled.
However, due to the three small singular values
the oblique projector along $\SWC$  
onto the whole subspace $\SV$
is very unstable, which fails to correctly separate the signals
in $\SV$ from the background. The result of applying 
the oblique projector onto the signal of the left hand graph is 
represented by the broken line in the right  hand graph.  
As can be observed, the projection
does not yield the required signal, which is
represented by the continuous dark line in the same graph.
Now, since the spectrum of singular values has a clear jump
(the last three singular values are far from the previous ones)
it might seem that one could regularize the  projection by truncation
of singular values. Nevertheless, such a methodology turns out to be 
inappropriate for the present problem, as it does not yield the correct
separation.   

Proposition \ref{tru1} below  analyzes the effect that regularization
by truncation of singular values has on  the resulting
projection.

\begin{figure}[!ht]
\label{f2}
\begin{center}
\includegraphics[width=8cm]{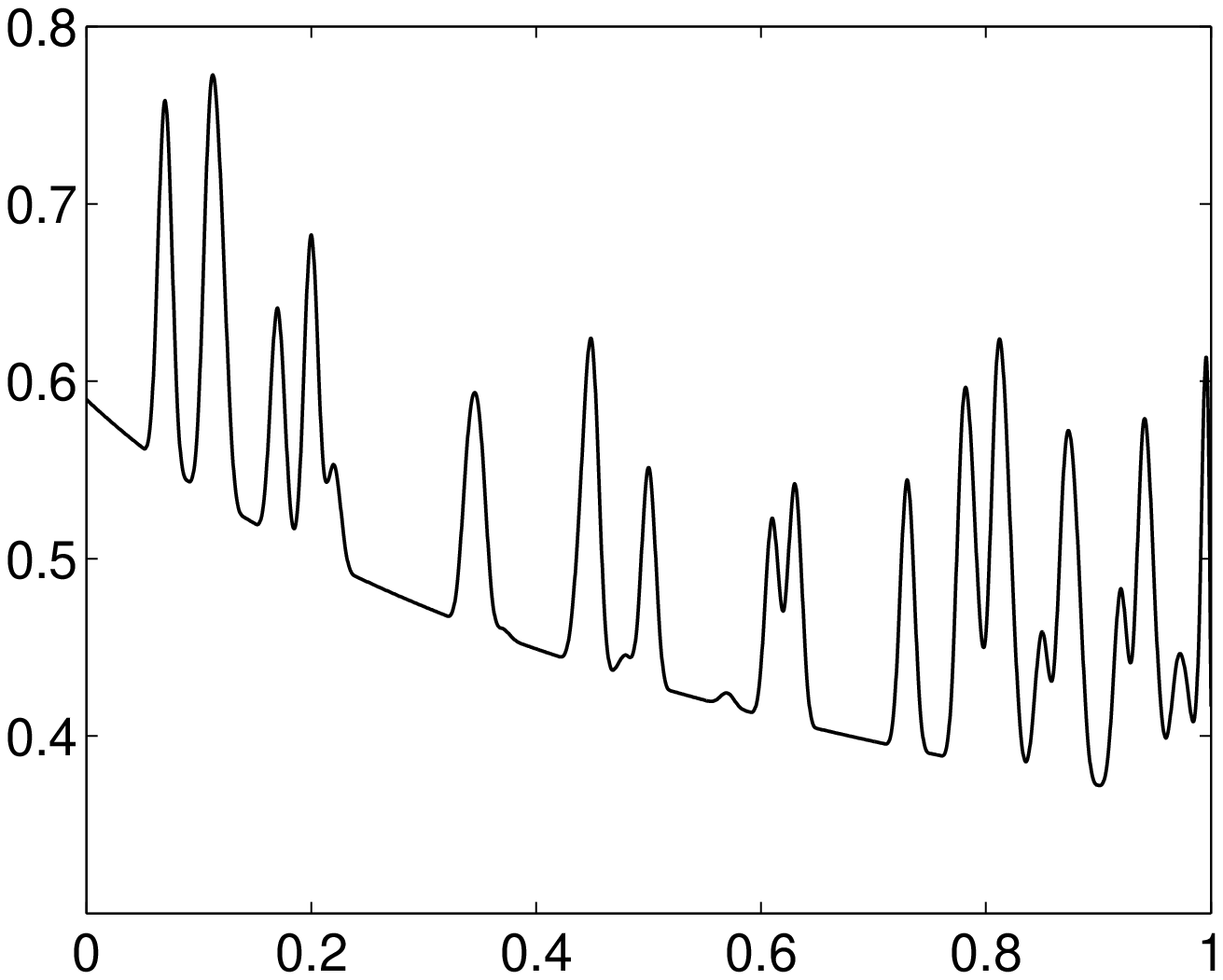}
\includegraphics[width=8cm]{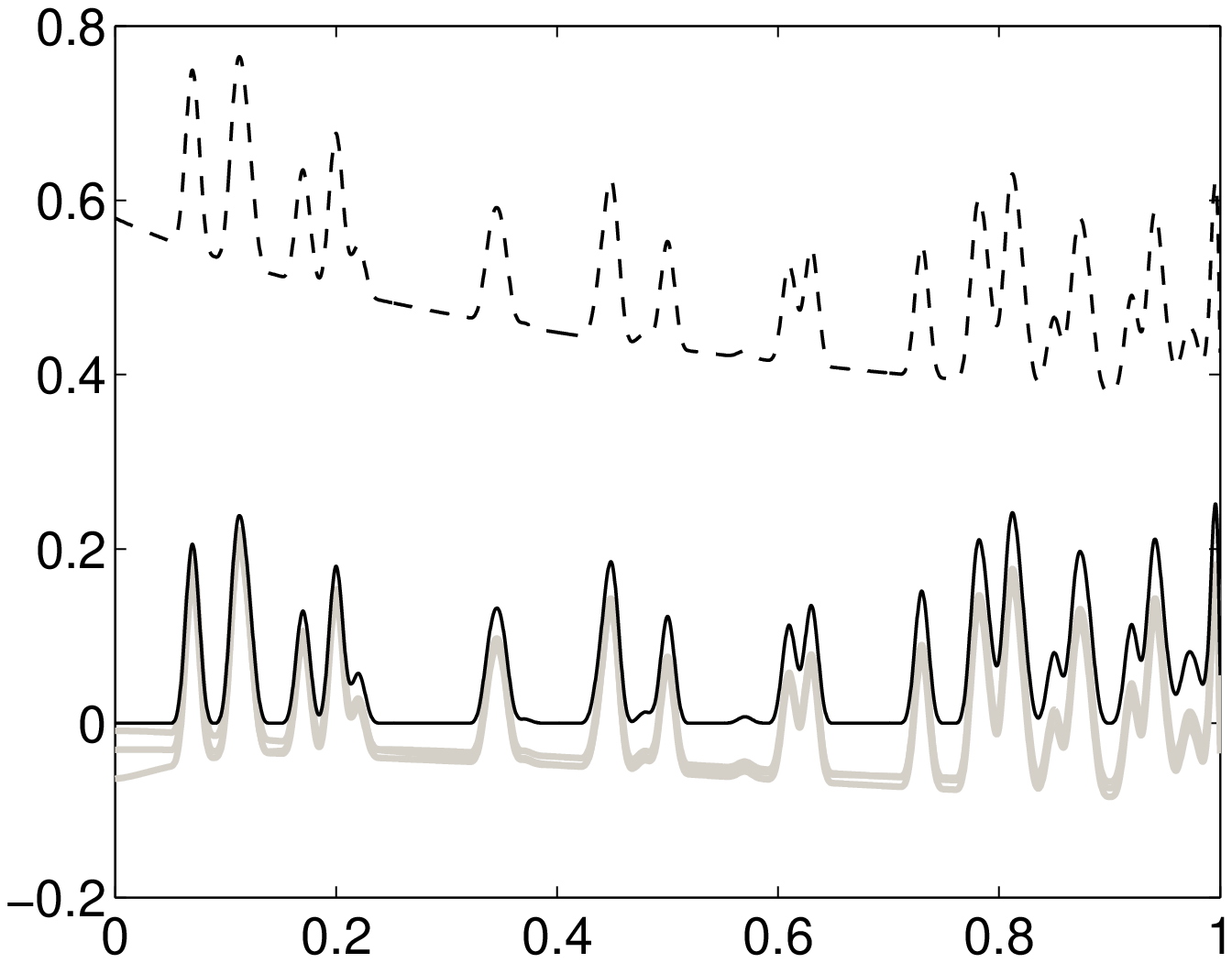}
\end{center}
\caption{\small{Left graph: signal plus background.
Right graph: the dark continuous line corresponds to the signal to be
discriminated from the one in the left graph. The broken line corresponds
to the approximation resulting from the oblique projection. The three close 
light lines correspond to the approximations obtained  by
truncation of one, two, and three singular values.}}
\end{figure}
\begin{proposition}
\label{tru1}
Truncation of the expansion \eqref{evw2} to consider up to $r$ terms,
produces an oblique projector
along $\SWr=\SWC + \SWo + \SVo$, with $\SVo= \Spann\{|\eta_i\ra\}_{i=r+1}^N$
and $\SWo= \Spann\{|\xi_i\ra\}_{i=r+1}^N$, onto $\SVr = \Spann\{|\eta_i\ra\}_{i=1}^r$. 
\end{proposition}
The proof of this proposition is to be found in Appendix B.

The above example illustrates very clearly the need for nonlinear
approaches. We know that a unique and stable solution does exist, since 
the signal which is to be discriminated from the background 
actually belongs to a subspace of the given spline space, and  
the construction of the oblique projectors onto 
such a subspace is well posed. However, the lack of knowledge 
about the subspace prevents us from separating the signal components 
by a linear operation. 
The greedy approaches that have been proposed for 
making tractable the search for the unknown subspace proceed in a 
stepwise manner \cite{Reb07b,Reb09}. Within those approaches, all the effort 
is focussed on the 
search for the right subspace for recursively constructing and 
adapting the vectors $|w_i\ra$ (c.f. \eqref{obli2}). 
Conceptually, the proposal in \cite{Reb09} implies to assume 
 a priori that {\em{none}} of the system states is populated and uses the 
 available signal to determine which  {\em {are}} the populated ones. Here 
 we wish to investigate the outcomes yielded by the converse 
 prior assumption, i.e., by considering a priori that all the 
 states are equally populated and use the available signal 
 to learn which  are the non-populated  ones. 

\section{The proposed nonlinear approach}
\label{sec4}
We start by recalling the available strategy for 
transforming the problem of discriminating the system's signal 
response $|f_\SV\ra$ from a given signal $| f \ra= |f_\SV\ra + |f_\SWC\ra$ into 
the problem of constructing the sparse representation of $|f_\SV\ra$ in 
$\SV$. 
Let us stress, once again that i)
the problem we need to face arises from the ill-posed feature 
of the oblique projectors onto the whole subspace $\SV$ and  ii) we 
work under the hypothesis that there exists an unknown
subspace ${\SV_K} = \Spann \{ {|v_{\ell_{i}}\ra}\}_{i=1}^K
\subset \SV$, where  $\{{\ell_{i}}\} _{i=1}^K$ is a set of $K$ unknown
indexes such that 
\be
\label{fsv}
|f_{\SV}\ra= |f_ {\SVKF}\ra =\sum_{i=1}^K c_{\ell_i} |v_{\ell_i}\ra.
\ee
Equivalently, \eqref{fsv} can be expressed in the form
\be
\label{fsu}
|f_{\SV}\ra= |f_ {\SVKF}\ra =\sum_{i=1}^M c_{i} |v_{i}\ra,\,\, \text{with}\,\,
c_{i}=0
\,\,\text{if}\,\, 
 i \neq {\ell_j},\,j=1\ldots,K.
\ee
Hence, to find the subspace ${\SV_K}$ is equivalent to finding
the sparse representation of $|f_{\SV}\ra$ in ${\SV}$, i.e.
a representation given by \eqref{fsu} where
only $K$ coefficients are nonzero. However we need to allow for the 
fact that we do not have access to the signal $|f_{\SV}\ra$
but only to the signal $|f\ra$. As proposed in \cite{Reb09}, 
we can deal somehow with this lack of information by noticing that
by applying the projector $\hat{P}_{\SW}$ both sides of \eqref{fsu} 
we have 
\be
\label{fsu2}
|f_\SW\ra = |f_{\SW_K}\ra= \sum_{i=1}^M   c_{i}  |u_i \ra,
\ee
where $|f_{\SW_K}\ra= \hat{P}_{\SW} |f_{\SVKF}\ra$ and $|f_\SW\ra= 
\op_{\SW} |f\ra$. 
Denoting  by $\hat{I}_{\SS}$ the identity operator in $\SS$ the 
projector $\op_{\SW}$ is obtained as 
$\op_{\SW} = \hat{I}_{\SS}- \op_{\SWC}$. Thus, since the subspaces
$\SS$ and $\SWC$ are known, we do have access to the component
$|f_{\SW}\ra$.  Because the coefficients in  \eqref{fsu} and 
\eqref{fsu2} are identical, one can find the sparse 
representation \eqref{fsu} by finding the sparse representation 
\eqref{fsu2} and using the resulting coefficients in \eqref{fsu}. 

At this point we begin to differ from the proposal in \cite{Reb09}. 
While in that publication the problem is tackled by a greedy stepwise search
for the indices  $\ell_i,\,i=1,\ldots,K$  in  
\eqref{fsv}, here we take a different route and  strive to 
find the sparse solution of \eqref{fsu2} by minimization of the 
$q-${\rm{norm}} like quantity
\be 
\label{qpnorm}
\| |c\ra \|_q^q= \sum_{i=1}^M |c_i|^q,\quad \text{with}\quad 0 < q \le 1.
\ee
The minimization of the $q$-norm$^q$ for 
determining a sparse solution has been studied   
in Mathematics and Signal Processing   
 and justified by the following consideration. The 
problem of finding the sparsest representation of a given model
is equivalent to
minimization of the zero norm $\|c\ra\|_0$ (or counting measure) 
which is defined as:
$$\||c\ra\|_0= \sum_{i=1}^M |c_i|^0$$ 
and therefore is equal to the number of nonzero entries of $|c\ra$. 
The minimization of $\||c\ra\|_0$ subject to linear
constraints is a classical problem of combinatorial search, which 
is in general NP-hard \cite{Nat95}. Thus,
 the minimization of $\sum_{i=1}^M |c_i|^q,$ for $ 0 < q \le 1$ has 
been considered \cite{RD99}. However since the minimization of  
   $\sum_{i=1}^M |c_i|^q,\, 
0 < q <1$ does not lead to a convex optimization problem, 
the most popular norm to minimize, 
 when a  sparse solution is required, is the 1-norm $\sum_{i=1}^M |c_i|$. 
 Minimization of the 1-norm is considered the best convex 
 approximant to the minimizer of $\||c\ra\|_0$ \cite{CDS98,CDS01}. Moreover,
it can be efficiently solved by linear programming techniques \cite{CDS98}. 
Since the problem of signal separation we are considering  
admits a unique solution, we are not particularly concerned about 
convexity. Hence we will set up our numerical strategy letting 
the parameter $q$ take any value in $(0,1]$. 

\subsection{Managing the constraints}

The optimization process we consider
is stated as follows:
{\em {Given the constraints \eqref{fsu2} minimize $\sum_{i=1}^M |c_i|^q.$}}

Now, in general, in order make use of constraints \eqref{fsu2} we need a 
numerical representation of $|f_\SW\ra$,  which in practice is
obtained by experimental measures. Thus, while restricting considerations 
to linear measurements we represent them as linear functionals, 
which, as established by Riesz' theorem \cite{RS80}, are amenable to  
representation by inner products  with some vectors. Accordingly,
we express  measures  on $|f\ra$ by the 
inner products 
$$m_j= \la m_j | f\ra,\quad j=1,\ldots,N.$$
The specification of the measurement vectors $|m_j\ra,\,j=1,\ldots,N$
  should be given in each particular case.  The 
 ones considered here have been chosen in relation to the
 examples we are presenting. Firstly, to simulate the observed 
 data we suppose that the measures are performed  by varying some 
 parameter (e. g. time, wavelength, temperature) that is 
 denoted as the variable `$x$'  discretized at the points 
 $x_j,\, j=1,\ldots,N$ to obtain the measures
 $$f_\SW(x_j)=\la x_j | f_\SW \ra,\quad j=1,\ldots,N$$ 
 and the corresponding linear functionals 
$$  \la x_j| u_i\ra= \la x_j| \hat{P}_{\SW}| v_i\ra,, \quad j=1,\ldots,N$$
from the state's signal responses $|v_i\ra,\,i=1,\ldots,M$. 
While the functionals $\la x_j | u_i\ra $ are modeled  
according to physical considerations, the values $f_\SW(x_j)$
are experimental data, thereby affected by errors. We then 
use the notation $f^o_\SW(x_j),\, j=1,\ldots,N$ to indicate the 
observations of $f_\SW(x_j),\, j=1,\ldots,N$. Consequently, 
rather than reproducing the data $f^o_\SW(x_j),\, j=1,\ldots,N$
we request that the model given by the r.h.s. of \eqref{fsu2}
satisfies the restriction
\be
\label{tol}
\sum_{j=1}^N (f^o_\SW(x_j) - f_\SW(x_j))^2 \le \delta,
\ee
$\delta$ accounting for the data's error.   
The stated optimization process subjected to 
this constraint is numerically difficult to realize. Nevertheless, 
we show here that the available information can  
be handled so as to successfully achieve the discrimination of 
 signal components, even when the data errors are
significant. For this we make use of an idea 
 we had introduced much earlier, in \cite{RCP93}, and 
 applied in \cite{RCP97}:  Replacing 
$f_\SW(x_j)$ by \eqref{fsu2}, the condition of minimal square distance
  $\sum_{j=1}^N (f^o_\SW(x_j) - f_\SW(x_j))^2$ 
 leads to the so called {\em{normal equations}}: 
\be
\label{noreq}
\la u_n| f_W^o\ra =\sum_{i=1}^M c_i \la u_n |u_i\ra,\quad n=1\ldots,M.
\ee
Of course, since we are concerned with ill posed problems
we cannot use all these equations to find the coefficients  
$c_i,\,i=1,\ldots,M$. However, as proposed in \cite{RCP93}, 
we could use `some'
of these equations as constraints of our optimization process. 
The number of such equations being the necessary to reach the condition 
\eqref{tol}. 

We have then transformed the original  problem into the one of minimizing 
 \eqref{qpnorm} subject to a number of 
equations selected from $\eqref{noreq}$, the
$\ell_n$-th$,\,n=1\ldots,r$ ones say. 
We leave for the moment the restrictions  
$c_i \ge 0,\,i=1,\ldots,M$. We should  
worry about them only if they were not satisfied. 

In line with \cite{RCP93} we select the subset of equations \eqref{noreq} 
in an iterative fashion.  We start by the initial estimation  
$c_i= C,\,i=1,\ldots,M$, where the constant $C$ is determined by
minimizing the distant between the model and the data. Thus,  
\be
C=\frac{\sum_{n=1}^M \la  u_n| f^o_\SW\ra}
{\sum_{i=1}^M \sum_{n=1}^M \la u_i | u_n\ra}.
\ee
With this initial estimation we `predict' the normal equations 
\eqref{noreq} and select as our first constraint the worst predicted 
by the initial solution, let this equation be the 
$\ell_1$-th one. We then minimize 
 \eqref{qpnorm} subject to the constraint
\be
\la u_{\ell_1} | f_\SW^o \ra
=\sum_{i=1}^M c_i \la u_{{\ell_1}} |u_i\ra,
\ee
and indicate the resulting coefficients as $c_i^{(1)},\,i=1,\ldots,M$. With 
these coefficients we predict equations \eqref{noreq} and select 
the worst predicted as a new constraint to obtain 
$c_i^{(2)},\,i=1,\ldots,M$ and so on. 
The iterative process is  stopped when the condition 
\eqref{tol} is reached. 

The reader may be aware that the proposed strategy involves highly nonlinear
equations and in many situations the number of
necessary constraints is large enough to generate  a troublesome
numerical task.  However, we have been able to solve the  simulation 
of the next section (comprising up to 57 constraints)
by recourse to the method for minimization of the $(q$-\rm{norm})$^q$
 published in 
\cite{RD99}. Such an iterative method, called FOCal Underdetermined 
 System Solver (FOCUSS) in that publication, 
is straightforward implementable. It 
evolves by computation of pseudoinverse matrices, which 
under the given hypothesis of our problem, and within 
our recursive strategy for feeding the constraints,  
are guaranteed to be numerically stable
(for a detailed explanation of the method see \cite{RD99}).

\subsection{Numerical Simulation} 
We test the proposed approach, first on the simulation of Example 1 
of Section 3, and 
then extend that simulation   
to consider a more realistic level of uncertainty in the data. 
Let us remark that the signal is meant to represent 
an emission spectrum consisting of the superposition of spectral lines 
(modeled by B-spline functions of support 0.04) which  are centered   
at the positions $(n-1)\Delta,\, n=0,\ldots,102$, with $\Delta = 0.01$. 
Since the errors in the data in Example 1 are not significant, 
the procedure outlined in the previous section  accurately
recovers the spectrum from the background, with any 
positive value of the $q$-parameter less than or equal  to  one. The 
result (coinciding with the theoretical one) is shown in the 
right hand top graph of Figure~3.

Now we transform the example into a more realistic situation by adding 
larger errors to the data. In this case, the data set is perturbed 
by Gaussian errors of variance up to $1\%$  of each data point. 
Such a piece of data is plotted in the left middle graph of Figure~3
and the spectrum extracted by the the proposed approach is 
represented by the broken line in the right middle graph of Figure~3, 
which is difficult to differentiate from the theoretical one (continuous 
line).

Finally we increase the data's error up to $3\%$ of each data point
(left bottom graph of Figure~3) and, in spite of the  perceived 
significant distortion of the signal, we could still recover
a spectrum which, as shown by the broken line 
in the right bottom graph of Figure~3, 
is a fairly good approximation of the true one (continuous line).
\begin{figure}[!ht]
\label{fig3}
\begin{center}
\includegraphics[width=7cm]{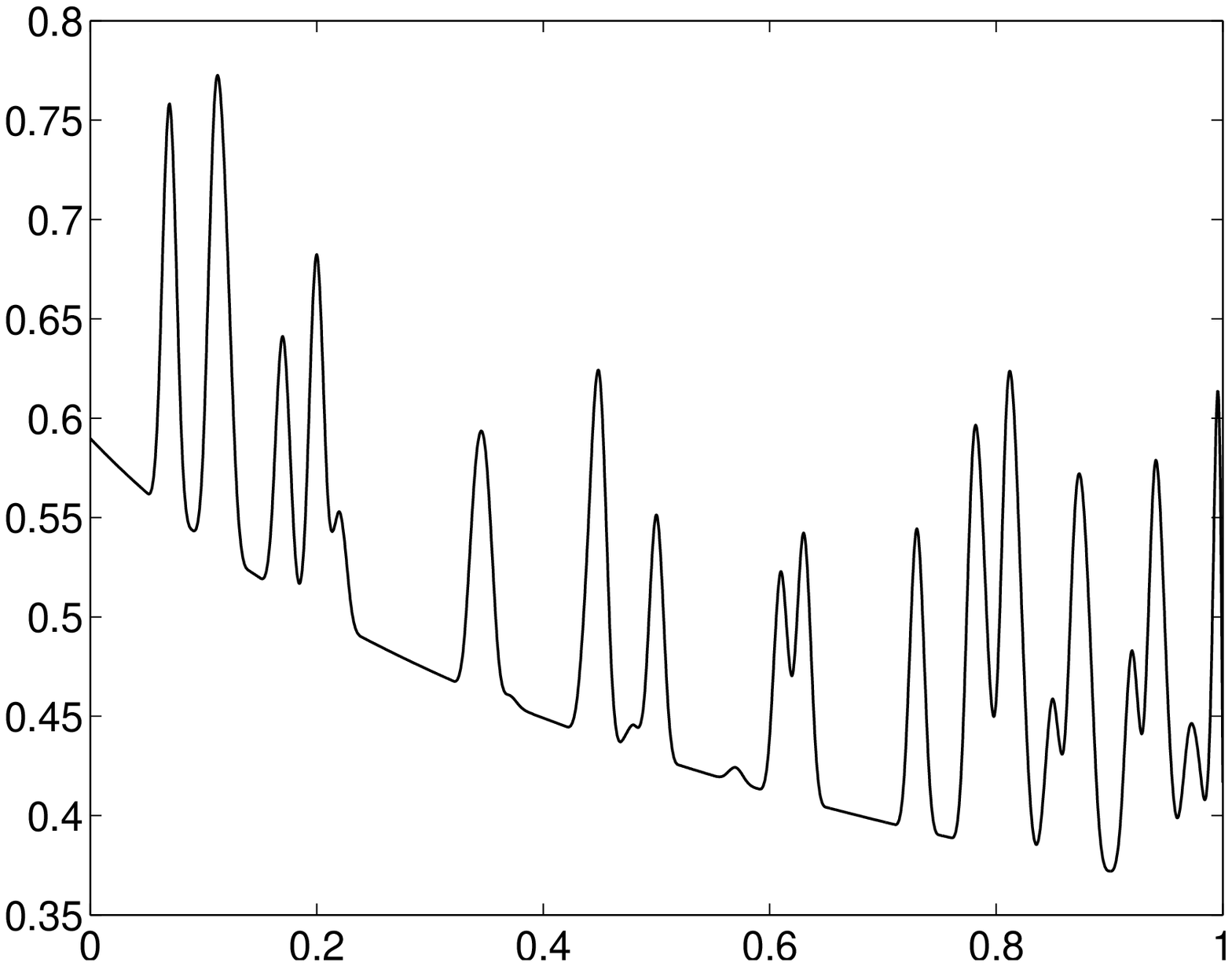}
\includegraphics[width=7cm]{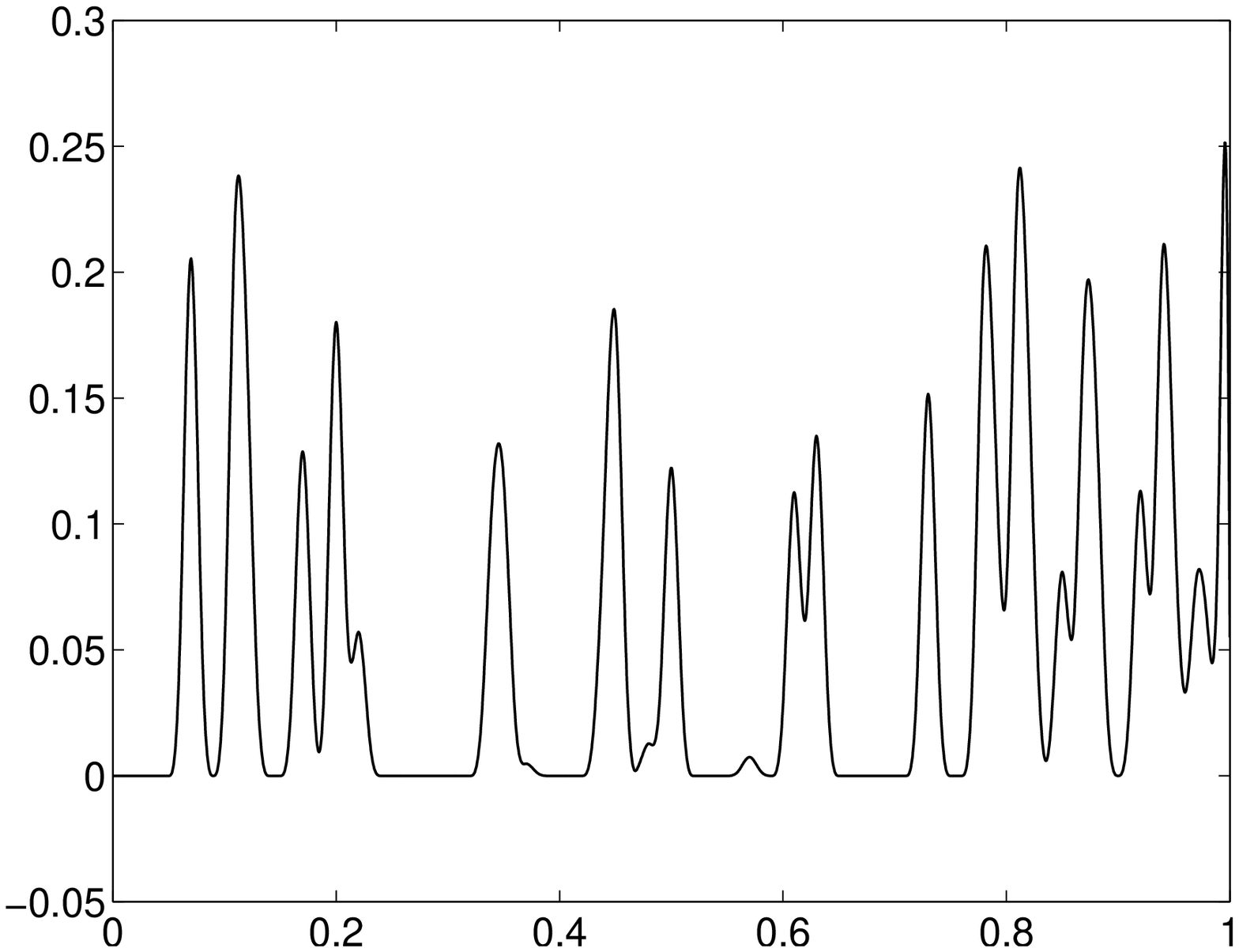}\\
\includegraphics[width=7cm]{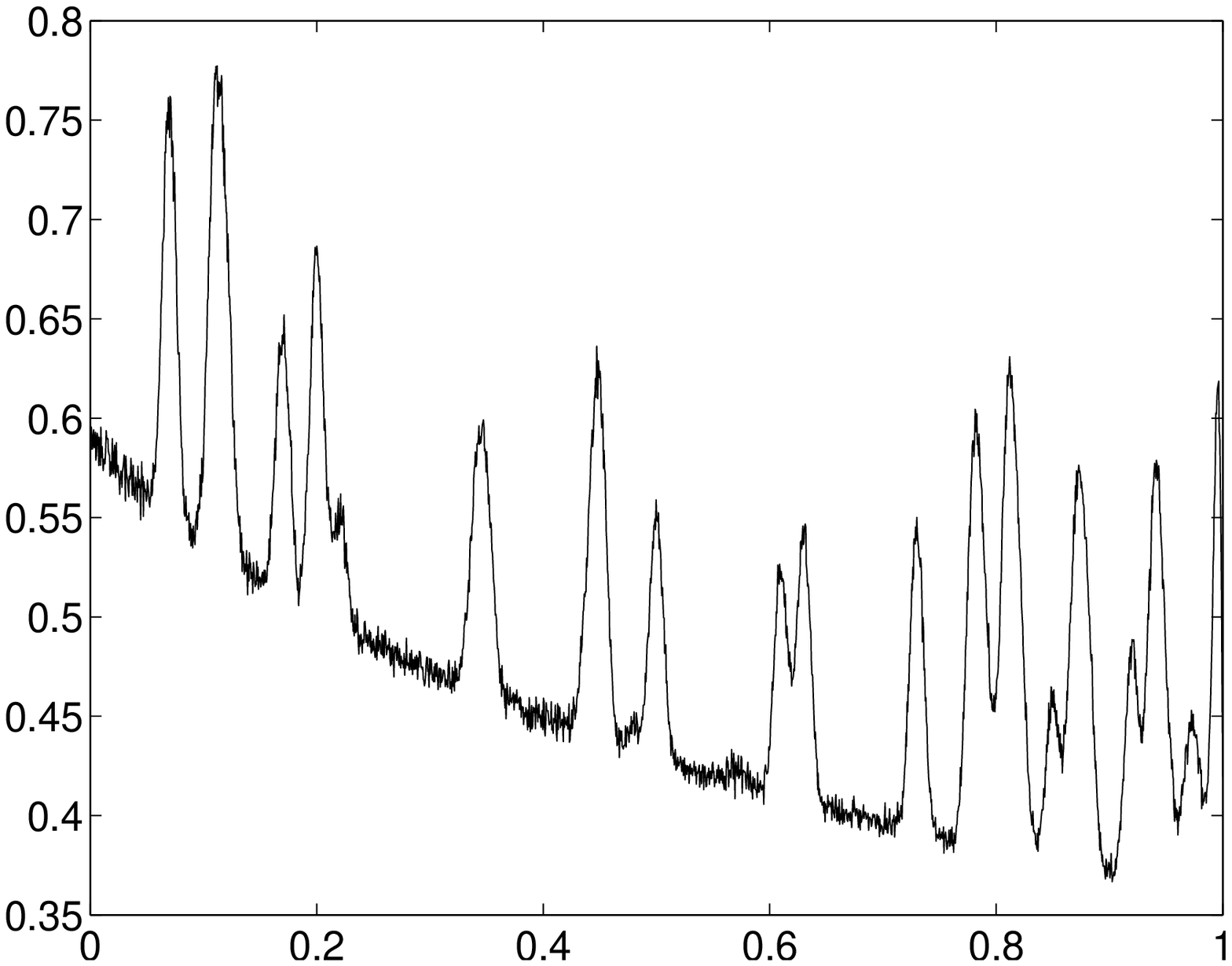}
\includegraphics[width=7cm]{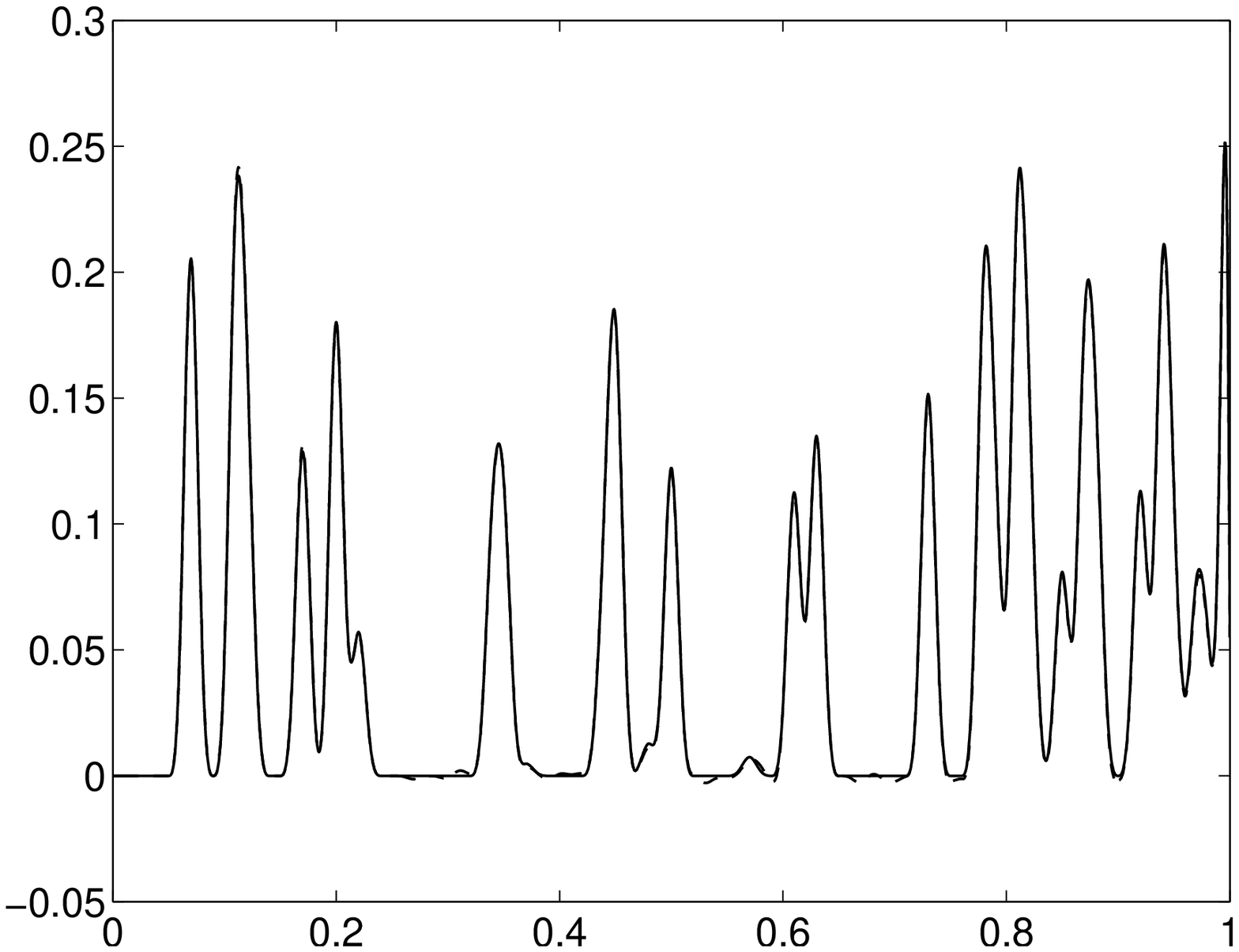}\\
\includegraphics[width=7cm]{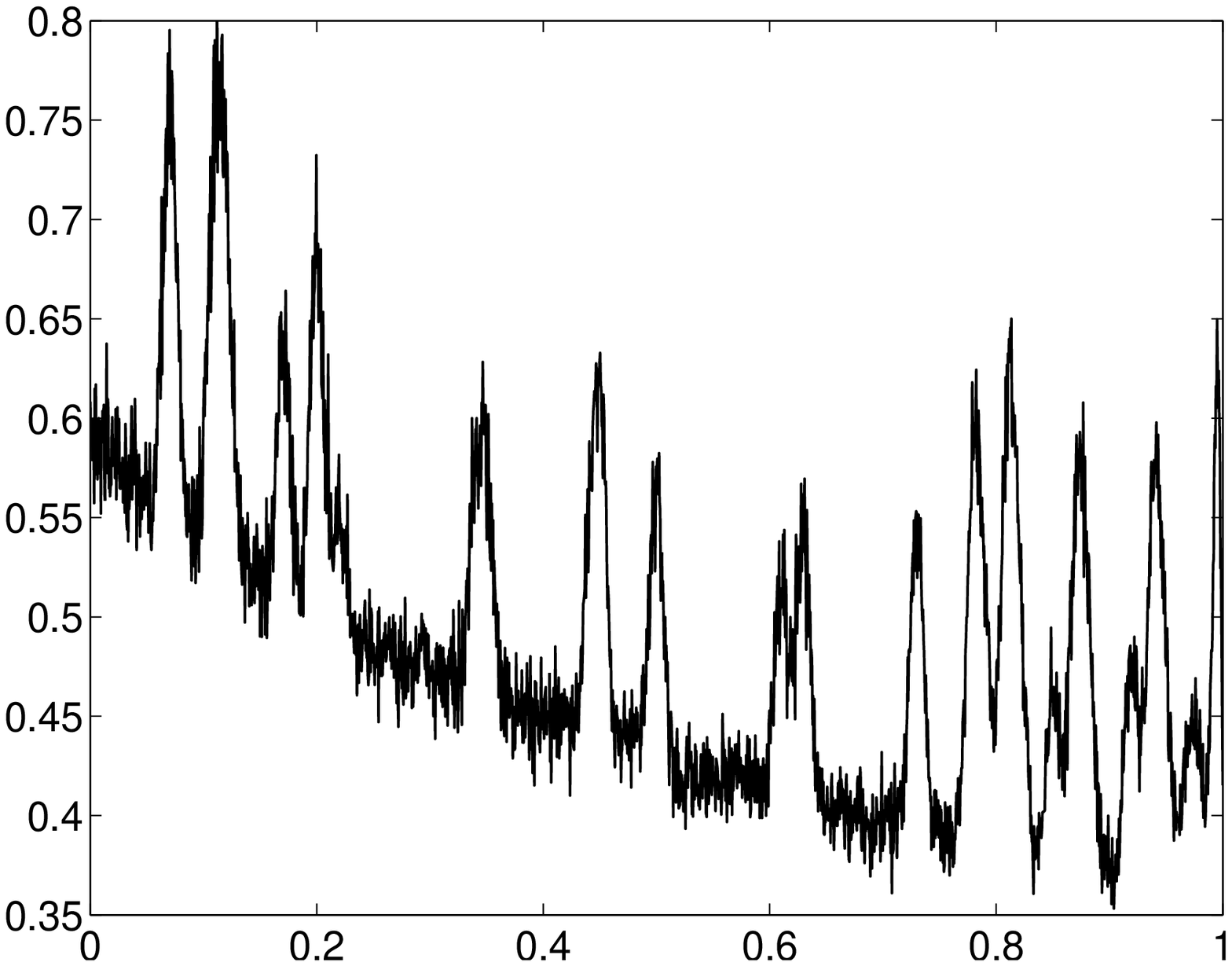}
\includegraphics[width=7cm]{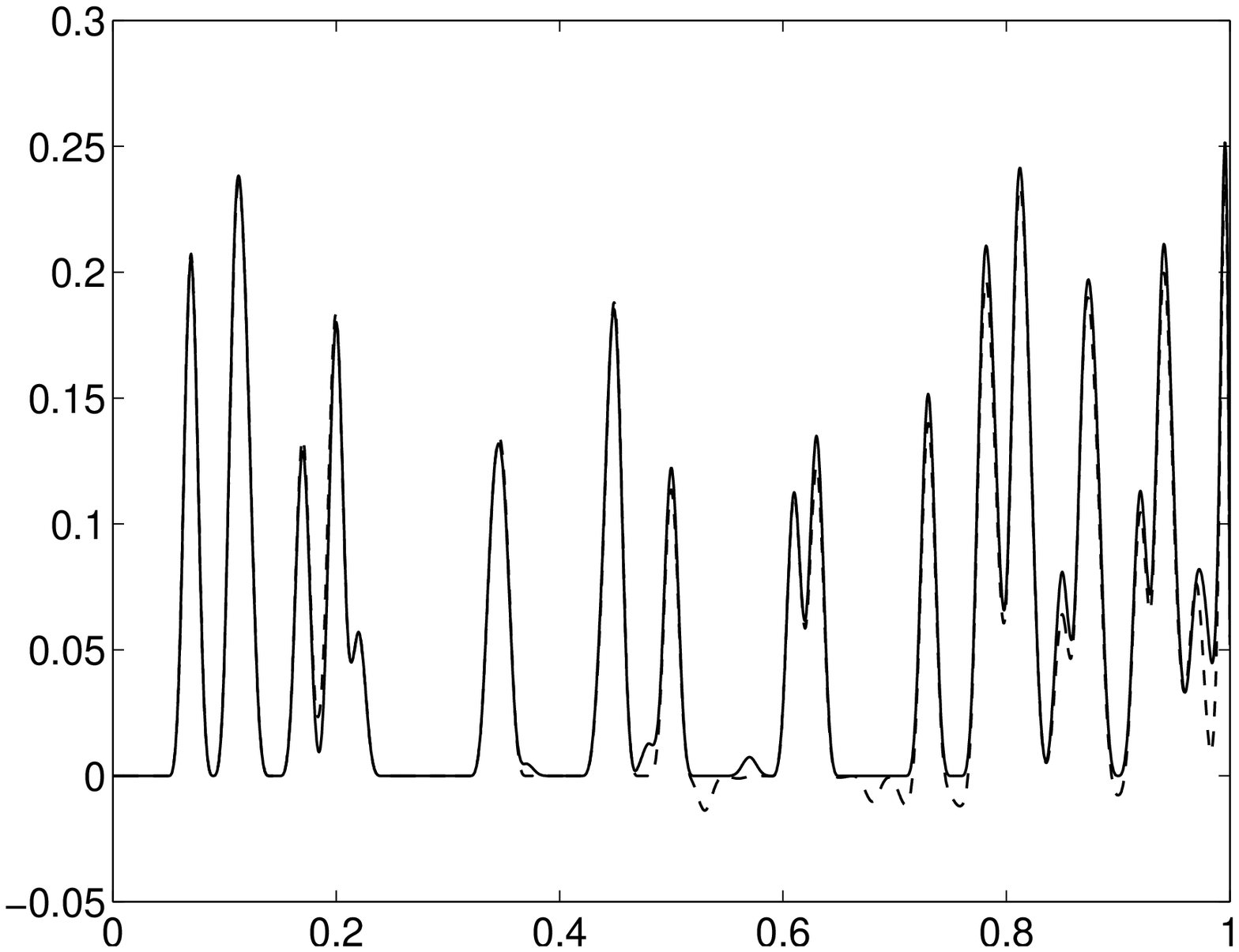}
\end{center}
\caption{\small{Top left graph: signal plus background generated as
described in Example 1 of  Section 3.
Top right graph: Recovered system' signal response,
which coincides with the true one.
Middle left graph: signal of Example 1 distorted  by
Gaussian errors of variance up to $1\%$ of each data value.
Middle right graph: the broken line represents the approximation
of the system' signal response
yielded by the proposed approach. The continuous line represents the
true signal.
Bottom graphs: Same description as in the previous graphs but
the data distorted by Gaussian errors of variance up to $3\%$ of
each data value.}}
\end{figure}
We have repeated the experiment for different realization of the 
errors (with the same variance) and the results remained essentially 
equivalent. Moreover, we have considered other realizations of the experiment
by drawing different spectra through the process described in 
Section 3.   
By observing the outcomes of a number the different realizations 
 we  can assert that the quality of the results shown in Figure~3 
is a fair representation of those obtained for different 
spectra. Variation of the $q$-value did not produce significant 
changes. The results of Figure~3 were obtained for the value 
$q=0.8$. The number of equations that were necessary to use 
in order to reach the stopping criterion  for  
the different level of error were:
$K=57$ for the data in the top graph, $K=51$ for the 
data in the middle graph and $K=43$ for the bottom graph.

It is appropriate to stress once more that for small level 
of errors the solution
of this example is unique. The numerical experiment illustrates
the fact that, for the degree of sparsity being considered  
(out to 103 states only 30 are populated) the solution can be 
reached by the whole range of $q$ values
in $(0,1]$. This is so because for all $q$ in the range $(0,1]$ 
the number of constraints which are needed to obtain the solution 
is still small enough to yield a well posed problem. 
Let us stress further that,  we have not
made explicit used of the constraints $c_i\ge 0,\,i=1,\ldots,M$  but look 
for the solution by  minimization of the quantity
$\sum_{i=1}^M |c_i|^q$, which is non-extensive for all $q$-values.
\section{Conclusions}
\label{conclu}
The problem of discriminating information produced by phenomena 
of different nature has been addressed through  a 
non-extensive nonlinear approach. The proposed framework is founded 
on the minimization of a $q-$norm like quantity. 
It is appropriate to remark that our main concern was to 
realize the  discrimination of information components in cases  
admitting a {\em {unique}} theoretical solution. The problem was 
transformed into an underdetermined linear one,  due to the 
numerical instability of the concomitant  full rank problem. 
The resulting approach has been tested by recourse to a numerical 
example which cannot be handled by linear techniques (even 
for unrealistically high quality data). 
A detailed analysis of the limitation 
affecting the linear technique has been provided. The 
nonlinear approach presented here was shown to be capable of
overcoming those limitations. It has  correctly
realized the required task, even for data distorted by 
significant random errors. We are aware that further studies 
 may  be in order and we are confident that the results presented 
here will motivate future works.

\section*{Acknowledgements}
Support from the Engineering and Physical Sciences Research Council (EPSRC),
UK, grant EP$/$D06263$/$1, is acknowledged.
\appendix
\subsection*{A. Proof of Proposition \ref{bio}}
\renewcommand{\theequation}{A.\arabic{equation}}
\setcounter{equation}{0}
\begin{proof}
Using \eqref{xi} and \eqref{eta} we have
$$
\la \xi_m |\eta_n\ra = 
 \frac{1}{\sigma_n \sigma_m}
\la\psi_n| \hat{U}^\ast\hat{V}|\psi_m\ra =\delta_{n,m}
\frac{\lambda_m}{\sigma_n \sigma_m}= \delta_{n,m},
$$
which proves the biorthogonality property.

The proof that $\Spann\{|\xi_n \ra\}_{n=1}^N=\SW$ stems from  the fact that
$\SW= \Spann\{|u_i \ra\}_{i=1}^M = \Spann\{|w_i \ra\}_{i=1}^M$,
which allows us to express an arbitrary
$|g \ra \in \SW$ as the linear
combination
$|g \ra=\sum_{i=1}^M a_i |w_i \ra$. Then, using \eqref{wdu2}, we have
$|g\ra= \sum_{n=1}^N \til{a}_n |\xi_n \ra$ with $\til{a}_n= \frac{1}{\sigma_n}
\sum_{i=1}^M a_i  \la i |\psi_n\ra$, which proves that
$\SW \subset \Spann \{|\xi_i \ra\}_{i=1}^N$. On the other hand
for $|g \ra\in \Spann \{|\xi_i \ra\}_{i=1}^N$ we can write 
$|g\ra= \sum_{n=1}^N d_n |\xi_n\ra$ and using \eqref{xi} 
we have $f \ra=\sum_{i=1}^M \til{d}_i |u_i \ra $,
with $\til{d}_i=  \frac{1}{\sigma_n}\sum_{n=1}^N d_n |\psi_n(i) \ra$.
This proves that $\Spann \{|\xi_i \ra\}_{i=1}^N \subset \SW$
and therefore $\Spann \{|\xi_n \ra\}_{n=1}^N=\SW$.
The proof that $\Spann \{|\eta_n \ra\}_{n=1}^N=\SV$ is equivalent
to the previous one.
\end{proof}
\subsection*{B. Proof of Proposition \ref{tru1}}
\renewcommand{\theequation}{B.\arabic{equation}}
\setcounter{equation}{0}
\begin{proof}
The biorthogonality between  $\{|\xi\ra\}_{i=1}^r $ and
$\{|\eta_i\ra\}_{i=1}^r$ established in
Proposition \ref{bio} ensures that
$\EVWr=\sum_{i=1}^r |\eta_i \ra \la \xi_i| $ is a projector, since
$\EVWr^2=\EVWr$.

As established in Proposition \ref{bio},
 $\SV=\Spann\{|\eta_i\ra\}_{i=1}^N$,
 and therefore every $|f\ra\in \SV$  can be
 decomposed as $|f\ra= |f_r\ra + |f_o\ra$ with $|f_r\ra\in \Spann\{|\eta_i\ra\}_{i=1}^r$ and
 $|f_o\ra \in \Spann\{|\eta_i\ra\}_{i=r+1}^N$. Moreover, $\EVWr |f\ra= |f_r\ra,
 \EVWr |f_r\ra= f_r\ra$, and $\EVWr |f_o\ra= 0$, which proves that the
 projection is onto $\SVr$ and $\SVo$ is included in the
 null space of $\EVWr$. Equivalently, for  every
 $|g_o \ra \in \SWo=\Spann\{|\xi_i\ra\}_{i=r+1}^N$ we have
 $\EVWr |g_o \ra=0$, because the set $\{|\xi_i\ra\}_{i=1}^N$ is orthonormal.
 Thus, $\SWo$ is included in the null space of $\EVWr$.
 \end{proof}

\newpage
\bibliographystyle{elsart-num}
\bibliography{revbib}

\begin{thebibliography}{10}
\expandafter\ifx\csname url\endcsname\relax
  \def\url#1{\texttt{#1}}\fi
\expandafter\ifx\csname urlprefix\endcsname\relax\def\urlprefix{URL }\fi

\bibitem{BS94}
R.~Behrens, L.~Scharf, Signal processing applications of oblique projection
  operators, IEEE Transactions on Signal Processing 42 (1994) 1413--1424.

\bibitem{Reb07a}
L.~Rebollo-Neira, Constructive updating/downdating of oblique projectors: a
  generalization of the Gram--Schmidt process, Journal of Physics A:
  Mathematical and Theoretical 40 (2007) 6381--6394.

\bibitem{Reb07b}
L.~Rebollo-Neira, Oblique matching pursuit, IEEE Signal Processing Letters
14~(10) (2007) 703--706.

\bibitem{Reb09}
L.~Rebollo-Neira, Measurements design and phenomena discrimination,
J. Phys. A: Math. Theor. 42 (2009) 165210.

\bibitem{Wic94} M. V. Wickerhauser, 
Adapted Wavelet Analysis from Theory to Software,  AK Peters, Ltd (2004).

\bibitem{RD99} B.D. Rao and K. Kreutz-Delgado,
An Affine Scaling Methodology for Best Basis Selection,
IEEE Trans. Sig. Proc. 47 (1999) 187--200.

\bibitem{Tsa88}
 C. Tsallis, Possible generalization of Boltzmann-Gibbs statistics,
  J. Stat. Phys., 52, (1988) 479.

\bibitem{Tsa09} C. Tsallis
Introduction to nonextensive statistical mechanics, 
Springer-Verlag, NY, (2009).


\bibitem{Pla1} A. R. Plastino, A. Plastino,  Tsallis
Stellar Polytropes and Tsallis' entropy, Physics Letters A,  174
(1993) 834--386.

\bibitem{Pla2}A. R. Plastino, A. Plastino,  Tsallis Entropy,
Erhenfest Theorem and Information Theory, Physics Letters A,  
177  (1993) 177--179.

\bibitem{LS00}B. R. La Cour, W. C. Schieve, 
Tsallis maximum entropy principle and the law of large numbers,
Phys. Rev. E 62 (2000)  7494--7496.


\bibitem{AO01} S. Abe, Y. Okamoto, 
Nonextensive Statistical Mechanics and Its Applications
Series: Lecture Notes in Physics , Vol. 560
Springer-Verlag, NY (2001).

\bibitem{Eld03}
Y.~Eldar, Sampling with arbitrary sampling and reconstruction spaces and
  oblique dual frame vectors, Journal of Fourier Analysis and Applications 9
  (2003) 77--96.

\bibitem{CMS05a}G. Corach, A. Maestripieri, D Stojanoff,
A Classification of Projectors,
Banach Center Publ. 67 (2005), 145--160.

\bibitem{CMS05b}G. Corach, A. Maestripieri, D Stojanoff,
Projections in operators ranges, Proc. Amer. Math. Soc. 134 (2005),
no 3, 765--788.

\bibitem{Sch81}
L.~Schumaker, Spline Functions: Basic Theory, Wiley, New York, 1981.

\bibitem{AR05}
M.~Andrle, L.~Rebollo-Neira, Cardinal {B}-spline dictionaries on a compact
  interval, Applied and Computational Harmonic Analysis 18 (2005) 336--346.

\bibitem{Nat95}
B.K. Natarajan, Sparse approximate solutions to linear systems,
SIAM J. Comput. 24 (1995) 227--234.



\bibitem{CDS98} S.S. Chen and D.L. Donoho and M.A. Saunders,
Atomic Decomposition by Basis Pursuit, SIAM Journal on Scientific Computing,
20 (1998) 33--61.

\bibitem{CDS01} S.S. Chen and D.L. Donoho and M.A. Saunders,
Atomic Decomposition by Basis Pursuit, SIAM Rev. 43 (2001) 129--156.

\bibitem{RS80}
M.~Reed, B.~Simon, Functional Analysis, Academic Press, London, 1980.

\bibitem{RCP93} L. Rebollo-Neira, A. Constantinides, A. Plastino, F. Zyserman, A. Alvarez,
 R. Bonetto, H. Viturro,
 Statistical inference, state distribution, and noisy data, 
 Physica A 198  (1993) 514--537

 \bibitem{RCP97} L. Rebollo-Neira, A. G. Constantinides, A. Plastino, 
 A. Alvarez,
 Bonetto, M. I\~niguez Rodriguez,
 Statistical analysis of a mixed-layer x-ray diffraction peak,
 Journal of Physics, 30, 17 (1997)
 2462--2469.

\end{thebibliography}
\end{document}